# Using PT-symmetry for switching applications

Anatole Lupu[1,2], Henri Benisty[3], Aloyse Degiron[1,2]

[1] Univ. Paris-Sud, Institut d'Electronique Fondamentale, UMR 8622, 91405 Orsay Cedex, France
[2] CNRS, Orsay, F-91405, France
[3] Laboratoire Charles Fabry, IOGS, CNRS, Univ. Paris-Sud, 2 Ave A. Fresnel, 91127 Palaiseau Cedex, France
*corresponding author, E-mail: `anatole.lupu@u-psud.fr`

**Abstract**

This work introduces a new class of PT-symmetry grating assisted devices for switching or modulation applications. Their operation is based on a four-wave interaction, thus marking a step in the further development of PT-symmetry devices which currently are essentially based on two-wave interactions. A remarkable feature of the new device is that all their properties also hold for the case of imperfect PT-symmetry operation, corresponding to the important practical case of fixed losses.

## 1. Introduction

The development of photonics enabled by the advent of nanofabrication technologies during the past decades has triggered the emergence of new types of artificial materials such as photonic crystals, metamaterials, plasmonic structures, and more recently so-called PT-symmetric devices [1-13], referring to Parity-Time symmetry. Aside from purely fundamental considerations, the tremendous interest for these synthetic materials is also strongly motivated by the practical outcomes targeting functionalities that can be achieved by gain/loss modulation in such structures. In recent contributions we showed that PT-symmetric couplers (PTSCs, obtained by coupling a lossy waveguide with another guide with gain) can operate as remarkably efficient switches when the proper amount of losses is included in the system [14-16].

In this contribution, we show that switching or modulation can also be achieved by considering a different type of PT-symmetric structures where gain/loss antisymmetry is implemented along the direction of light propagation. In the spirit of our previous work [14-16], we demonstrate that operation compatible with practical applications can be achieved with an imperfect PT-symmetry design, where only the gain is variable while the losses are set to a fixed value.

## 2. PT-symmetry Bragg grating assisted directional coupler switch

The operation principle of PT-symmetric couplers, explored for switching applications in [14-16], is based on the modification of propagation constants achieved by gain/loss modulation. In these structures based on a transverse modulation of gain and loss, the switching operation exploits the characteristic eigenvalue diagram of PT symmetric systems on one side only of the symmetry-breaking exceptional point. The aim of the present study is to show that the same gain/loss modulation principle can be used for switching and routing applications in the case of longitudinal type PT symmetry structures where the complex index profile with antisymmetric imaginary parts is built along the direction of light propagation [17-19]. An issue tackled only scarcely up to now is the nature and magnitude of the opportunities offered by such structures. Since a four-port device is required for switching or routing applications, we consider the example of a passive uniform waveguide coupled with a PT-symmetric Bragg grating (BG). It can be implemented either in a vertical or horizontal type coupling geometry. This provides an additional degree of freedom for device engineering since multiplexing properties shall differ with the coupling geometry.

A sketch of a PT-symmetric Bragg grating (BG) assisted vertical directional coupler is shown in Fig. 1(a). Its practical realization can be envisioned for example by means of hybrid integration technology of III-V/SOI semiconductors [20-22]. The vertical coupling geometry offers the possibility of using different types of materials, thus bringing an additional degree of freedom in engineering the waveguide dispersion properties.

The operation principle of the considered devices is based on the modification of the dispersion properties induced by the introduction of the PT-symmetry through the gain/loss level modulation in the grating structure. As is known, a periodic modulation of the real part of a grating waveguide effective index leads to the formation of a stop-band around the Bragg wavelength [top graph Fig. 2(a)]. By engineering the dispersion properties of the uniform waveguide, it is possible to tune its effective index branch so as to span the area inside the forbidden stop-band of the BG assisted waveguide [Fig. 2(b)] [23,24]. The phase matching between the waveguides is then frustrated and coupling between them remains negligible [23-27].

By bringing periodically modulated gain and loss in the BG waveguide, it is possible to build a nearly PT-symmetric complex coupling profile. The variation of the imaginary part of the coupling profile in such a PT-symmetric BG is quarter-period-shifted with respect to the variation of its real part [middle graph on Fig. 2(a)]. The amplitude of the variation for the real and imaginary parts of the coupling profile is the same as in the perfectly PT-symmetric case. The characteristic feature of the PT-symmetry Bragg grating (PTSBG), the one-guide/two-port device, is that the forbidden stop-band and its associated dispersion disappear. In the case of PTSBG the dispersion relation is akin to that of a uniform waveguide [17-19]. The phase becomes matched and phase matching occurs at the intersection point of the passive uniform and PTSBG dispersion curves [Fig. 2(b)].

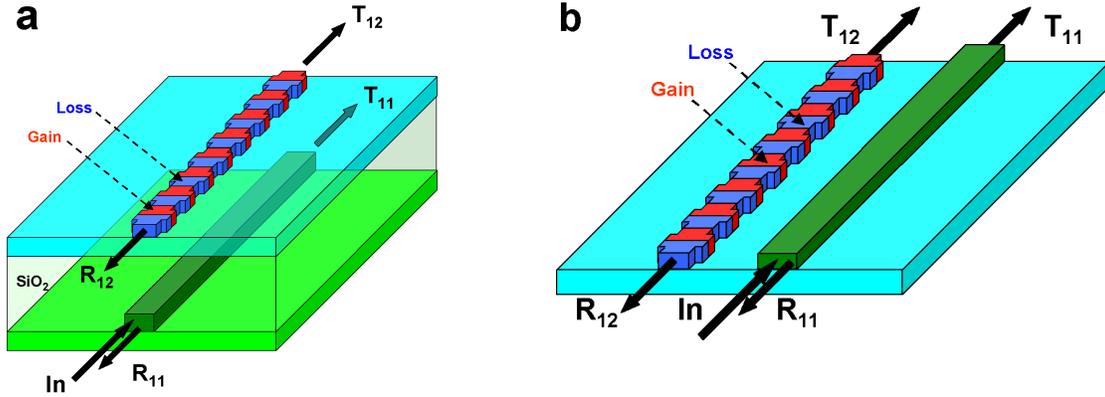

Fig. 1: Sketch of a PT-symmetry BG assisted directional coupler with rectangular grating profile. Gain and loss grating sections are indicated by red and blue colors, respectively. a) Vertical coupling geometry; b) Horizontal coupling geometry.

The implementation of simultaneous gain and loss modulation required for a perfect PT-symmetric case is however not easy. For a practical realization, it is much more convenient to vary the gain only while fixing the losses to a constant value. This situation is illustrated by the bottom graph in Fig. 2(a) which corresponds to the case of a modulation of loss-only (i.e. not gain) still in quadrature combined with the modulation of the real index BG. It turns out that the dispersion characteristic of such an imperfect PT-symmetric BG is essentially similar to that of a passive BG. The stop-band, though spectrally narrower, is still present and phase matching is frustrated. Consequently, as shown in Fig. 4(a), there is no notable coupling or back-reflection due to the BG waveguide when the light is injected into the uniform waveguide.

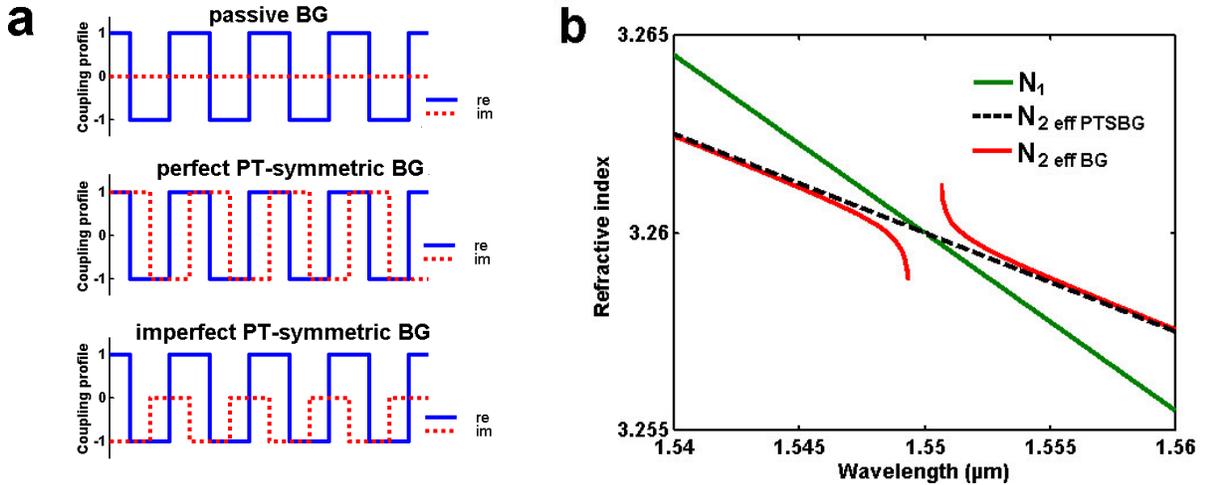

Fig. 2: a) Real (blue solid) and imaginary (red dotted) parts of the grating complex coupling profile. Top: without gain and loss modulation. Middle: equal gain/loss and index modulation. Bottom: without gain, loss modulation is half of index modulation; b) Effective index of the uniform waveguide (green) and the conventional BG assisted waveguide (red) and PT-symmetry BG (dashed).

By bringing gain, the system becomes perfectly PT symmetric, the gap disappears and phase matching occurs. As evidenced in Fig. 4(b), this leads to an efficient energy transfer between the two waveguides in the vicinity of the phase matching wavelength. The modification of the BG dispersion can thus be used for the implementation of switching or add-drop operation.

The operation principle of the horizontal type directional coupler is basically similar to that of the vertical type coupler. The main difference is that in the absence of grating, the dispersion curves of two waveguides are generally identical [dashed curve in Fig. 2(b)]. In the absence of gain, with only loss and real index being modulated, and corresponding to the case of imperfect PT-symmetry, a stop-band is formed in the vicinity of the Bragg wavelength. As a consequence, coupling between the waveguides is forbidden in the stop-band region but occurs for the rest of the spectral range [Fig. 5(a)].

By bringing gain in the system, the condition for perfect PT-symmetry is fulfilled. The dispersion curves of two waveguides coincide and coupling is allowed for all wavelengths [Fig. 5(b)]. In contrast, for the vertical type coupler, frustration extends over the whole spectral range and phase matching occurs only for a limited spectral interval in the vicinity of the Bragg wavelength). The operation of the two types of couplers are thus complementary and provide additional flexibility for switching and add-drop operation.

## 3. Four-waves coupled modes approach modeling

A coupled mode theory (CMT) approach has been used to verify the assertions inferred from the phase matching analysis of Section 2 and to explore the PTSBG directional coupler behavior in detail. For this purpose, the investigated device was schematized as shown in Fig. 3. The schematized device consists of a five-layer structure (from $N_1$ to $N_5$) with two parallel slab waveguides surrounded by claddings. The waveguide in the upper part bears a double side BG of $s$ periods with a total length L=$s\Lambda$. Both waveguides are assumed to be single mode. The dashed lines traced for the BG assisted waveguide indicate the original waveguide width without grating modulation. This width is in turn used to determine the effective index and the propagation constant $\beta_j=2\pi n_j/\lambda$, which is introduced in the system of coupled mode equations.

A rectangular grating profile is assumed in such a way that the coupled-mode equations lead to analytical solutions. Following the approach developed in [26-30], the device is decomposed along the propagation axis $z$ in a series of parallel waveguide segments of lengths $\Lambda^{++}$, $\Lambda^{-}$, $\Lambda^{+}$, $\Lambda^{-+}$ delimited by the grating corrugations and gain/loss sections. Since all the coupling matrices are independent of $z$ within each section, the coupled-mode equations can be solved exactly. To define the elements of the coupling matrices, we use an approach based on individual waveguide modes. The detailed description of the method can be found in [31].

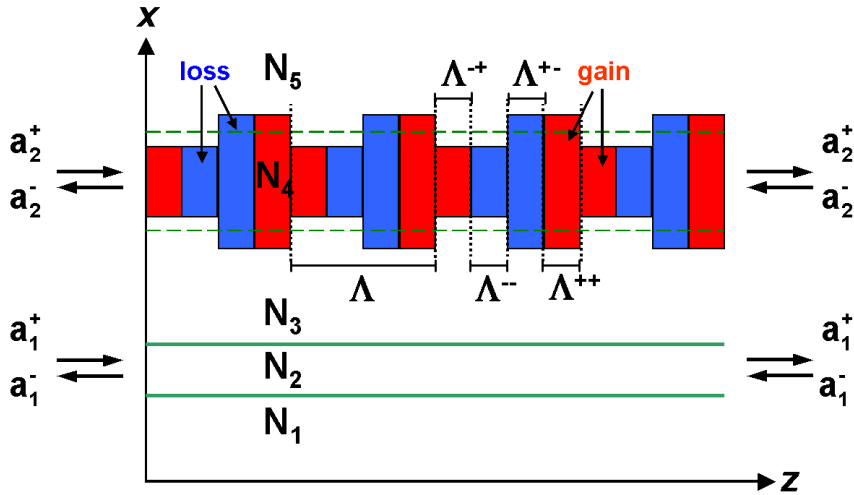

Fig. 3. Schematic representation of the PTSBG directional coupler with rectangular grating profile.

The relation governing the exchange between forward and backward propagating waves of two coupled waveguides on the grating sections is:

$$\frac{d}{dz}A = jM_{pq}A \quad (1)$$

where $A$ is the column vector $A=[a_1^+, a_1^-, a_2^+, a_2^-]^T$ with field amplitude elements and $M_{pq}$ are $4\times 4$ matrices with constant value coefficients:

$$M_{pq} = \begin{pmatrix} \beta_1 & p\chi_1' + iq\chi_1'' & \kappa & p\chi_{12}' + iq\chi_{12}'' \\ -(p\chi_1' + iq\chi_1'') & -\beta_1 & -(p\chi_{12}' + iq\chi_{12}'') & -\kappa \\ \kappa & p\chi_{12}' + iq\chi_{12}'' & \beta_2 & p\chi_2' + iq\chi_2'' \\ -(p\chi_{12}' + iq\chi_{12}'') & -\kappa & -(p\chi_2' + iq\chi_2'') & -\beta_2 \end{pmatrix} \quad (2)$$

Here $\beta_1$, $\beta_2$ are the waveguide propagation constants, $\chi_1'$, $\chi_1''$ and $\chi_2'$, $\chi_2''$ are the real and imaginary part of the BG coupling coefficient provided here for the sake of generality of Eq. (2). In the actual case, $\chi_1'=0$, $\chi_1''=0$ since only the second waveguide is bearing a grating. For a rectangular grating profile, the relation between the BG coupling coefficient $\chi_2$ and the waveguide effective index modulation $\Delta n_2$ is:

$$\chi_2 = \frac{\Delta n_2}{\Lambda n_2} \quad (3)$$

The terms $\kappa$ and $\chi_{12}$ stand for the co-directional evanescent-coupling and contra-directional Bragg exchange evanescent-coupling, respectively:

$$\kappa = \frac{\pi}{2L} \quad (4)$$

$$\chi_{12} = \Lambda(\chi_1 + \chi_2)\kappa \quad (5)$$

Finally $p=\pm1$ and $q=\pm1$ account for the real and imaginary part index profile modulation for different grating sections $\Lambda_{pq}$.

It is important to note that while transmission is identical, irrespective to the choice of initial light input waveguide, the contrary is true for reflection. When changing the light injection from the front-side to the end-side, the reflection properties are substantially different. A negligibly small direct or cross reflection is observed for the case shown in Figs. 4(b), 5(b) and an amplified reflection in the spectral region corresponding to the BG stop-band is observed for the case shown in Figs. 4(c), 5(c). The reflection non-reciprocity is essentially similar to that of a PTSBG, but in our case it occurs for both direct and coupling mediated reflections. It is interesting to note that in all cases the direct reflection $R_{11}$ is higher than its coupling mediated counterpart $R_{12}$. Such a behavior is totally counterintuitive when light injection is performed into the uniform waveguide that neither bear gain/loss or refractive index modulation.

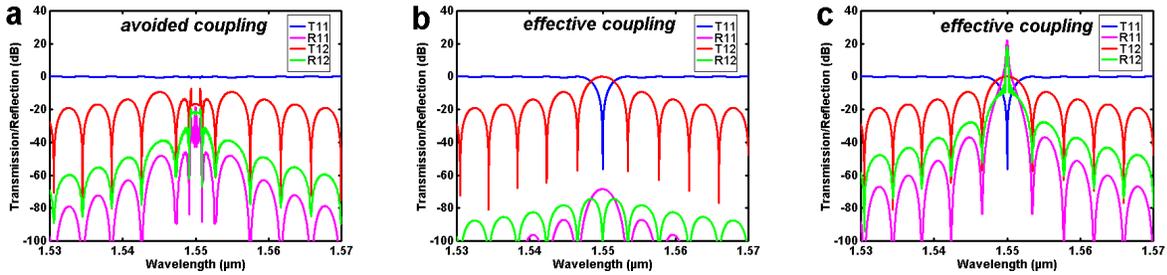

Fig. 4: Vertical coupling geometry. a) Frustrated coupling regime spectral response; b) & c) Phase matching allowed PTSBG operation differing by the choice of the light injection from the front-side or end-side, respectively.

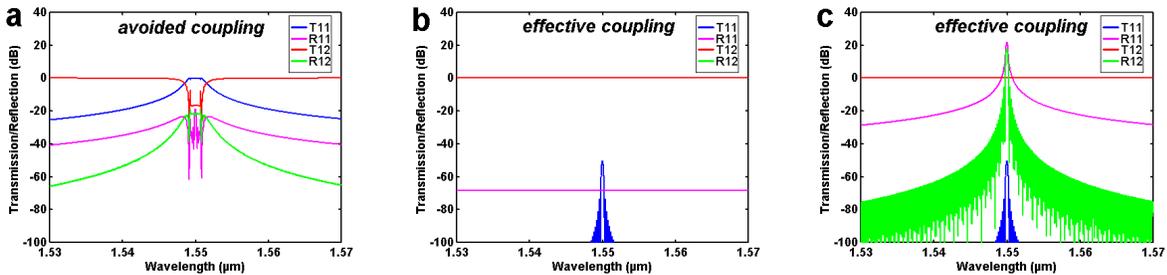

Fig. 5: Horizontal coupling geometry. a) Frustrated coupling regime spectral response; b) & c) Phase matching allowed PTSBG operation differing by the choice of the light injection from the front-side or end-side, respectively.

This result means that the description of the device behavior cannot be reduced to a framework of two-wave interactions and that it is crucial to account for all four interacting waves. This point marks an important difference with respect to the previously reported works on PT-symmetry grating assisted devices. Even though some of them are treating devices involving a four-wave interaction, the analysis of their properties is still based on a conventional two-waves approach [33].

It could be noted that some asymmetry in reflection is also observed for the case of imperfect PT-symmetric BG assisted directional couplers whose results are shown in Figs. 3(a), 4(a). The reflection level is however generally quite small for both front and end input configurations and can be neglected.

## 4. Conclusions

We propose a new strategy to implement switching or routing in PT-symmetric couplers. A very interesting aspect of our approach is that all the described behavior and properties also hold for the case of imperfect PT-symmetry operation, corresponding to the important practical case of fixed losses. By a proper engineering of the complex index profile, it is still possible to maintain the frustrated-coupling regime in the absence of gain. As explained, this corresponds to the forbidden phase matching. By bringing gain into the system, perfect PT-symmetry is restored and so is the phase matching condition. Switching or add-drop operations are thus achieved from the sole gain variation, which is much more convenient for practical applications. The calculated examples shown in Figs. 4(a-c) and 5(a-c) correspond to this case of fixed losses and variable gain operation mode, with indices typical of III-V materials. The requested tuning ranges for switching being of the same order of magnitude as those typical in optoelectronics, practical realizations exhibiting such frustrated or favored coupling behaviors for further assessment seem at hand.